\documentclass{PoS}

\title{Recent progress in the development of large area\\silica aerogel for use as RICH radiator in the\\Belle II experiment}

\ShortTitle{Large area silica aerogel for use as RICH radiator in the Belle II experiment}

\author{\speaker{Makoto Tabata}$^{,a,b}$, Ichiro Adachi$^c$, Hideyuki Kawai$^b$, Shohei Nishida$^c$ and Takayuki Sumiyoshi$^d$\\
\llap{$^a$}Institute of Space and Astronautical Science (ISAS), Japan Aerospace Exploration Agency (JAXA), Sagamihara, Japan\\
\llap{$^b$}Department of Physics, Chiba University, Chiba, Japan\\
\llap{$^c$}Institute of Particle and Nuclear Studies (IPNS), High Energy Accelerator Research Organization (KEK), Tsukuba, Japan\\
\llap{$^d$}Department of Physics, Tokyo Metropolitan University, Hachioji, Japan\\
        E-mail: \email{makoto@hepburn.s.chiba-u.ac.jp}}

\abstract{We report recent progress in the development of large-area hydrophobic silica aerogels for use as radiators in the aerogel-based ring-imaging Cherenkov (A-RICH) counter to be installed in the forward end cap of the Belle II detector, which is currently being upgraded at the High Energy Accelerator Research Organization (KEK), Japan. The production of approximately 450 aerogel tiles with refractive indices of either 1.045 or 1.055 was completed in May, 2014, and the tiles are now undergoing optical characterization. Installation of the aerogels was tested by installing them into a partial mock-up of the support structure.}

\FullConference{Technology and Instrumentation in Particle Physics 2014,\\
		2--6 June, 2014\\
		Amsterdam, the Netherlands}

\begin{document}

\section{Introduction}

A proximity-focusing aerogel-based ring-imaging Cherenkov (A-RICH) counter is installed in the forward end cap of the Belle II detector \cite{cite1}, which is currently being upgraded at the High Energy Accelerator Research Organization (KEK), Japan. The A-RICH counter uses silica aerogel Cherenkov radiators with a refractive index ($n$) of approximately 1.05 to identify charged pions and kaons and has a separation capability greater than 4$\sigma $ at momenta up to 4 GeV/$c$. When filling the large (3.5 m$^2$) end-cap region with aerogel radiators, it is necessary to minimize the number of aerogel tiles (i.e.  maximize the aerogel dimensions) to reduce tile boundaries, because the number of detected photoelectrons decreases at the boundaries. To achieve this goal, we developed a method to produce large-area transparent aerogel tiles with no cracks.

For our purposes, aerogels may be produced in one of the two ways, i.e. by the modernized conventional method \cite{cite2} or by the pin-drying \cite{cite3} method. With the pin-drying method, more transparent aerogels can be produced than with the conventional method. However, in view of the result of prototypes, we decided to use the conventional method to mass produce large-area aerogel tiles for the detector \cite{cite4}, because this method proved to be more cost-effective and provided a high yield of large-area crack-free aerogel tiles.

We next developed an aerogel-radiator tiling scheme \cite{cite4} for the end cap of the Belle II detector. The cylindrical end cap will be filled with 124 segmented dual-layer-focusing aerogel combinations (248 tiles in all) based on a multilayer-focusing radiator scheme \cite{cite5}. This tiling scheme calls for 18 $\times $ 18 $\times $ 2 cm$^3$ large-area aerogel tiles with $n$ = 1.045 and 1.055 for the upstream and downstream layers, respectively. The aerogel tiles with different refractive indices were manufactured separately and will be stacked during the installation stage. With a water jet cutter, the aerogel tiles will be trimmed in a fan shape to fit the cylindrical end-cap geometry (concentric layers 1 to 4, counting from the center of the end cap), making the best use of the hydrophobic features. In an electron-beam test performed at the Deutsches Elektronen-Synchrotron (DESY), we verified that the $K$--$\pi$ separation capability of a prototype A-RICH counter exceeded 4$\sigma $ at 4 GeV/$c$, where we used a prototype focusing combination aerogel tile trimmed with the water jet cutter.

\section{Status of mass production and optical characterization}

\begin{figure}[t] 
\centering 
\includegraphics[width=0.53\textwidth,keepaspectratio]{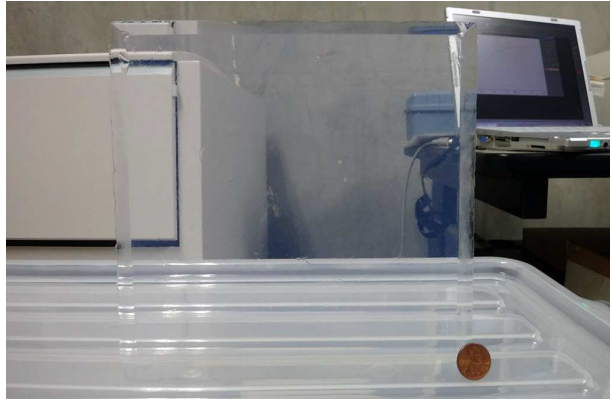}
\caption{First aerogel sample delivered. The refractive index and transmission length were 1.0444 and 56~mm, respectively. The tile measured 18 $\times $ 18 $\times $ 2 cm$^3$.}
\label{fig:fig1}
\end{figure}

From September, 2013 to May, 2014, the Japan Fine Ceramics Center collaborating with Mohri Oil Mill Co., Ltd mass produced the aerogels for the actual A-RICH counter. Since November 2013, a total of 449 tiles have been delivered to KEK. Figure \ref{fig:fig1} shows the first aerogel sample that was delivered. In parallel, we characterized the aerogels by visually checking them and measuring their refractive index, transmission length  and density using the methods described in Ref. \cite{cite2}. At the end of May 2014, 239 tiles have been optically characterized.

\begin{figure}[t] 
\centering 
\includegraphics[width=0.51\textwidth,keepaspectratio]{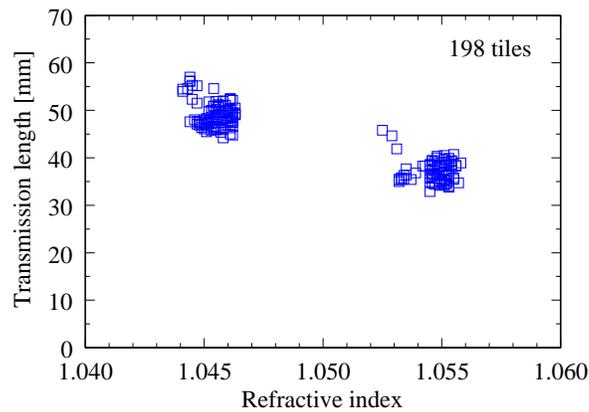}
\caption{Transmission length at 400 nm as a function of refractive index. The refractive index was measured with a 405-nm semiconductor laser.}
\label{fig:fig2}
\end{figure}

To date, we have confirmed 95 (70) tiles with $n$ = 1.045 ($n$ = 1.055) as good samples (i.e. candidates for installation into the actual detector). The crack-free yield after supercritical drying was 91\% (outperforming our target of 80\%), and 69\% of the 239 tiles were undamaged and fulfilled our transparency requirements. After characterizing all the aerogel tiles delivered, we should have over 300 good ones. Figure \ref{fig:fig2} shows the results of the optical measurements. The modernized conventional method led to aerogel tiles with well-controlled refractive indices and optimal transmission lengths. Note that once aerogels are trimmed by the water jet, we cannot remeasure their refractive index because the water jet machined surface strongly scatters the laser beam used in the measurement. Finally, we determined that the weight (density) of the tiles was useful to distinguish the upstream and downstream aerogels, in addition to management by an identification number.

\section{Mock-up installation test}

In March, 2014, we tested the water jet method of trimming the mass-produced aerogel tiles. Tatsumi Kakou Co., Ltd. machined a total of eight tiles (four tiles for each refractive index) with a water jet cutter. The aerogels were trimmed to fit the shape of the second and third concentric layers. Dimensioning errors were less than 0.5\% compared with the design (true size was smaller than design size in most cases).

\begin{figure}[t] 
\centering 
\includegraphics[width=0.53\textwidth,keepaspectratio]{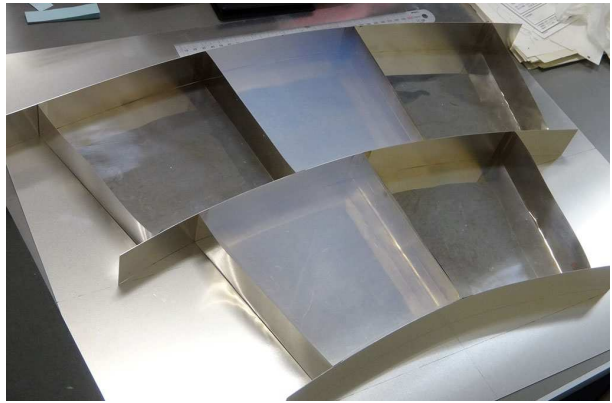}
\caption{Partial mock-up of aerogel support structure. Water-jet-trimmed two-layer aerogel combinations were installed in the bottom left box (second concentric layer) and in the upper center box (third concentric layer).}
\label{fig:fig3}
\end{figure}

In April, 2014, we tested the installation of the aerogel tiles. A partial mock-up of the aerogel support structure (made of 0.1- to 0.5-mm-thick aluminium sheets) was prepared to hold the aerogel tiles on edge in the end cap (Figure \ref{fig:fig3}). The mock-up consisted of five boxes shaped to match the second and third concentric layers in which the aerogel tiles will be installed. One by one, we carefully placed the water-jet-trimmed aerogel tiles in the boxes. The installation was successful because there was sufficient margin between the aerogel tiles and the boxes.

\section{Conclusion}

We developed large-area, transparent aerogel tiles for use as radiators in the A-RICH counter of the Belle II experiment. For the actual detector, the aerogel tiles were mass produced and are now being optically characterized. In addition, we prepared a partial mock-up of the aerogel support structure, in which we installed several aerogel tiles.

\section*{Acknowledgements}

The authors are grateful to the members of the Belle II A-RICH group for their assistance. We are also grateful to the Japan Fine Ceramics Center, Mohri Oil Mill Co., Ltd. and Tatsumi Kakou Co., Ltd. for their contributions to mass producing the aerogel tiles and water jet machining. This study was partially supported by a Grant-in-Aid for Scientific Research (A) (No. 24244035) from the Japan Society for the Promotion of Science (JSPS). M. Tabata was supported in part by the Space Plasma Laboratory at ISAS, JAXA.


\begin{thebibliography}{99}
\bibitem{cite1} 
T. Abe \emph{et al.}, \emph{Belle II technical design report}, \emph{KEK Rep.} 2010-1, 2010 [{\tt 1011.0352}].

\bibitem{cite2} 
M. Tabata \emph{et al.}, \emph{Hydrophobic silica aerogel production at KEK}, \emph{NIMA} {\bf 668} (2012) 64 [{\tt 1112.3121}].

\bibitem{cite3} 
M. Tabata \emph{et al.}, \emph{Recent progress in silica aerogel Cherenkov radiator}, \emph{Phys. Proc.} {\bf 37} (2012) 642 [{\tt 1203.4060}].

\bibitem{cite4} 
M. Tabata \emph{et al.}, \emph{Silica aerogel radiator for use in the A-RICH system utilized in the Belle II experiment}, \emph{NIMA}, (2014) in press [{\tt 1406.4564}].

\bibitem{cite5} 
T. Iijima \emph{et al.}, \emph{A novel type of proximity focusing RICH counter with multiple refractive index aerogel radiator}, \emph{NIMA} {\bf 548} (2005) 383 [{\tt physics/0504220}].

\end{thebibliography}
\end{document}